\documentstyle[11pt,paspconf,epsf]{article}
\def\hmpc{\; {\rm h}^{-1}\; {\rm Mpc}}
\def\etal{et al.~}

\begin{document}

\title{Redshift Surveys and Large--Scale Structure}

\author{L. Guzzo}
\affil{Osservatorio Astronomico di Brera, Via Bianchi 46, I--22055, Merate
(LC), Italy}





\begin{abstract}

I review the present status of the mapping of the large--scale
structure of the Universe through wide--angle redshift surveys.  
In the first part of the paper, I discuss
the current state of the art, describing in some detail the recently 
completed ESO Slice Project.  In the second, I recall some of the basic 
motivations for performing larger surveys.  Then, in the final 
section, I present in detail the three large redshift survey projects that 
in the years from now to the end of the Century are expected to satisfy 
this demand.

\end{abstract}




\section{Introduction}

In preparring my contribution for this meeting, I thought it was a good 
moment to stop and to try and summarize the presently fervent activity 
in mapping the large--scale distribution of luminous matter, with a 
special look at the large redshift projects which are currently in their early
phases.  In fact, during these last twenty years since when 
Chincarini \& Rood (1976) recognized the ``segregation of redshifts'' 
in their maps of the  surroundings of the Coma cluster (actually the 
embryo of the following CfA surveys), the industry of redshift 
measurements has skyrocketed. 
As a consequence, our knowledge of, at least, the local (cz $<10000$ 
km s$^{-1}$) Universe has greatly improved (see Rood 1988 for a colourful
account of the pioneering years; more recent reviews can be found
in Giovanelli \& Haynes 1991, and Strauss \& Willick 1995).
Still, all the surveys 
within this depth show inhomogeneities of comparable size (e.g. the 
Great Wall in the North Galactic Cap, Geller \& Huchra 1989, or 
the Perseus--Pisces filament in the South, Giovanelli \etal 1986), making 
it difficult to think of them as statistically representative.  

The development of wide--field multi--object (MOS) spectrographs using
fibre optic advanced technology represents the key for the further 
giant step in the mapping of large--scale structure.
The first two wide--angle surveys using fibre spectrographs,
the ESO Slice Project and the Las Campanas Redshift Survey, have 
been recently completed, pushing the depth of the samples
to six times that of CfA2 (section~\ref{maps}).  However, the need
for larger, more three-dimensional maps is still strong, as I
shall briefly outline in section~\ref{stat}  There are three large
mapping projects which are expected to fill this gap, two using galaxies
and one using clusters of galaxies.  These, the Sloan Digital Sky Survey,
the 2dF survey and the ESO Key--Programme survey of ROSAT clusters will
be discussed in some detail in the last section.

\section{Cosmic Maps: the Present Picture} \label{maps}

In terms of redshift surveys we can now talk of the ``local'' Universe
as that mapped by the CfA2 and SSRS redshift surveys, to a depth of 
$\sim 100 \hmpc$.  I will not discuss here the many important results 
obtained from this combined data set (for which details can be found,
e.g., in Park \etal 1994), but prefer to concentrate on the results 
from the first large {\sl multiplexed} surveys.  I will also not discuss
here all the series of redshift surveys based on the IRAS catalogue
(treated exhaustively by Rowan--Robinson in this same volume),
and sparse surveys like the APM/Stromlo (Loveday \etal 1992).

By multiplexing, one means one of the major technological advances 
in the field of astronomical spectroscopy, i.e. the development of 
{\sl Multiple--Object} spectrographs.  These give simply the 
possibility of coupling more than one slit or aperture
in the focal plane of the telescope to the spectrograph, collecting
many spectra at once.  This is 
performed either through redirection of the light beam by means 
of optical fibres, or through slitlets carved out of a metal mask
placed in the focal plane.   While the latter instruments (e.g. EFOSC
at the ESO 3.6~m telescope) have their field of view limited by the
size of the detector, typically a few arcminutes, fibre spectrographs 
can exploit the whole corrected field of the telescope, with 
diameters of the order of 1 degree (see e.g. the review by Hill 1988).

The impact of multiplexing on large--scale redshift work is readily understood
by considering as an example the case of the ESO Optopus system, with 50 
fibres over a 30 
arcmin field at the Cassegrain focus of the ESO 3.6~m telescope.  The 
density of fibres on the sky for this instrument is then $\sim250$ 
deg$^{-2}$.  If we consider the number counts law for an euclidean geometry, 
\begin{equation}
N(<m) = N_\circ\;10^{0.6m}
\end{equation}
one can then ask for which value of $m$ the mean number counts will match
the density of fibres, and find that (e.g. in the IIIaJ blue band) this
happens for $m_{b_J}\sim 19-19.5$.  The simple introduction of a similar
instrument, therefore, immediately pushes the potentially optimal 
magnitude limit to four
magnitudes deeper than the CfA2.

\subsection{The ESO Slice Project}

The ESO Slice Project (ESP) is a first example of a large--scale, 
wide--angle galaxy redshift survey performed by mapping the sky 
through a fibre spectrograph, and actually the very first of such
surveys using one of the ESO telescopes.  The ESP started in 1991, 
after a number of unsuccesfull attempts and in a somewhat reduced 
form with respect to the original idea. It covers a $32^{\circ} 
\times 1^{\circ}$ strip at 
constant declination, (with a $5^\circ$ gap), centered at $\delta -40^{\circ}$,
using the Optopus fibre spectrograph at the 3.6~m ESO telescope.  The strip
is mapped with a regular grid of Optopus fields,
observing all the galaxies with $b_J \leq 19.4$ in the Edinburgh--Durham 
Southern Galaxy Catalogue (EDSGC, Heydon--Dumbleton et al. 1989).  
This magnitude limit, which optimizes the fibre coupler, results in an 
effective depth of $\sim 600\hmpc$.
The optical fibres cover 2.4 arcsec on the sky, and are manually plugged into 
a pre--drilled aluminum plate.  The data are now fully reduced, and the 
final redshift catalogue contains 3348 galaxies.  The overall redshift 
yield of the survey has been larger than 80\%.  More details 
can be found in Zucca \etal (1996).

\begin{figure}
\vspace{1.75in}
\plotfiddle{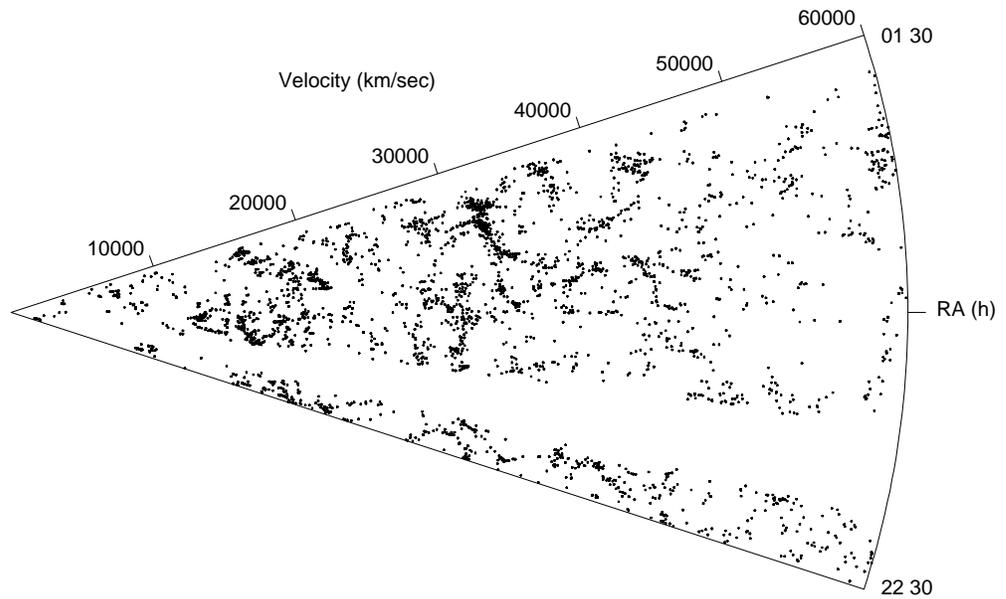}{2truecm}{0}{60}{60}{-190}{-20}
\caption{The ESO Slice Project data, for $cz\le 60000\; km\;s^{-1}$.  
The slice is $1^\circ$ thick, comprised between -39$^\circ$~45$^\prime$ and 
-40$^\circ$~45$^\prime$ in declination.} \label{esp-cone}  
\end{figure}

Fig.~\ref{esp-cone} shows a wedge diagram of the whole ESP data.
The first qualitative impression when comparing this plot to, e.g.,
the CfA survey slices, is that there does not seem to be any structure
with dimensions comparable to the survey size. Are we finally approaching
a fair sample of the Universe?

A rather large fraction of galaxies in the survey ($\sim 50\%$) show
emission lines in their spectra ([OII] $\lambda 3727$, $H\beta\; \lambda 4861$,
[OIII] $\lambda 4959$ and $\lambda 5007$).  These objects could be either 
spiral galaxies, where lines originate mostly from HII regions in the 
disk, or starburst galaxies.  Their distribution is different from that 
of galaxies without emission lines, a fact that could either 
be a manifestation  
of the known morphology--density relation, or indicate that starburst 
phenomena occur preferentially in low density environment, or both.

\begin{figure}
\vspace{1.75in}
\plotfiddle{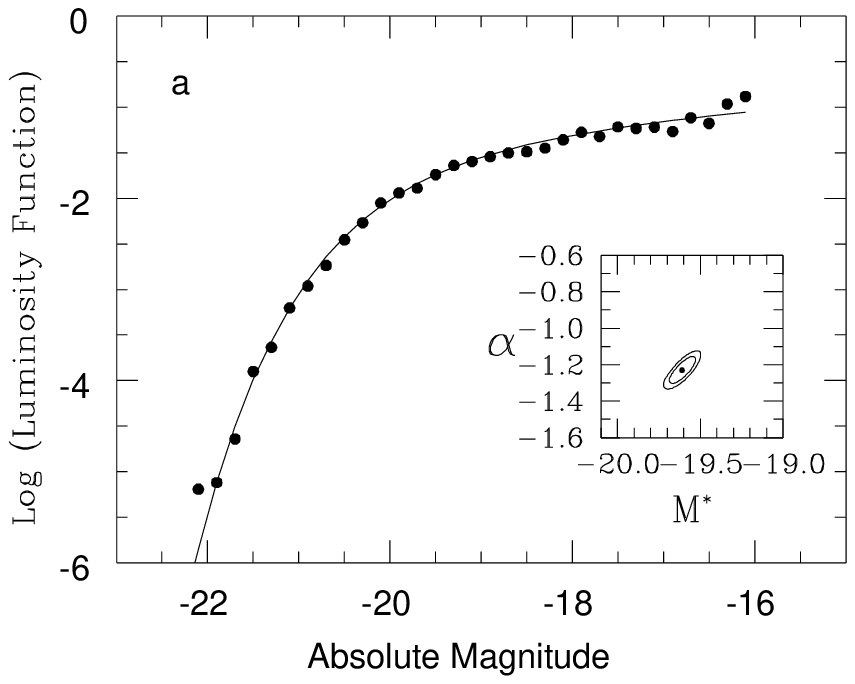}{1truecm}{0}{75}{75}{-325}{-292}
\plotfiddle{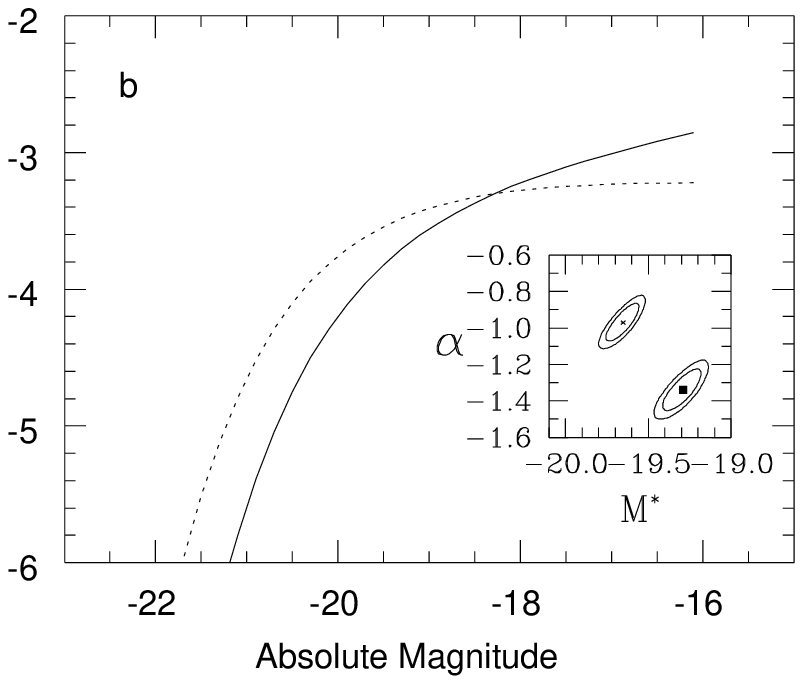}{0truecm}{0}{75}{75}{-128}{-120}
\caption{a) The ESP luminosity function for $M_{b_J} < -16 $,
estimated using a modified version of the non--parametric C--method
(Lynden--Bell 1971).  The inset shows the maximum--likelihood 68\% and 95\% 
confidence ellipses on $\alpha$ and M$^*$.  b) Normalized 
Schechter fits to the luminosity functions for ESP galaxies with (solid line) 
and without (dotted line) emission--lines.  Note the 
very significant separation in parameter space of the two data sets.
(From Vettolani \etal 1996).} \label{esp-lf}
\end{figure}

Most of the effort in the analysis of the new ESP data has been devoted 
so far to the study of the luminosity function (LF).  This has involved 
a careful study of the K--correction.   In particular, since galaxy 
morphology, a necessary ingredient to the K--correction, is not discernible 
for most of the galaxies in the survey, a statistical approach was adopted
as discussed in Zucca \etal (1996).  The Schechter function is a good fit
to the data for $M_{b_J}<-16$, as shown in fig.~\ref{esp-lf}a. At fainter 
luminosities (down to $-12$), a statistically significant raise above the 
Schechter form is detected (see Vettolani \etal 1996).   The best estimate
of the Schechter shape parameters, based on 3311 galaxies and using the 
maximum likelihood method of Sandage \etal (1979) (with arbitrary 
normalization), gives $\alpha = -1.23 $ and $M^*_{b_J} = -19.61 $.  
An interesting result is obtained when the LF is estimated separately
for galaxies with and without emission lines, as shown in fig.~\ref{esp-lf}b.
The difference in the parameters is very significant, with the population
of line--emitting objects rising steeply towards fainter magnitudes.


\subsection{The Las Campanas Redshift Survey} 

The other large multiplexed redshift survey recently completed is the 
Las Campanas Redshift Survey (LCRS).  It has been performed by the 
so--called KOSS collaboration (see e.g. Shectman 1995), using
a 112--fibre spectrograph with a 1.5$^\circ$ diameter field at the 
Cassegrain focus of the Du Pont 2.5~m telescope.  This is at present 
the largest survey available, with nearly 25000
redshifts.  Similar in depth to the ESP, but selected in the red band 
from CCD exposures, it covers a larger area -- about 700 deg$^2$ -- 
distributed among 6 strips.   There is not enough space here to detail
further on this project.   The reader is referred to
the contribution by Landy, in this same volume, and to references therein.
In particular, see Tucker (1994) for a comprehensive discussion of the sample
construction, and Lin (1995) for an account of various statistical analyses. 

\section{Quantitative Needs for Larger Surveys} \label{stat}

I have deliberately selected two scientific issues which contain in
my view the main motivations for larger redshift surveys.  I will 
discuss them here in a very schematic way.

\subsection{Correlation Function and Power Spectrum}

The correlation function $\xi(r)$ and its Fourier transform, the power
spectrum $P(k)$ are some of the simplest, but most important
clustering statistics (e.g. Peebles 1980).  Estimates of $P(k)$ from
present redshift surveys (e.g. Park \etal 1994), or large angular 
catalogues like the APM survey (Baugh \& Efstathiou 1993) are limited
to wavelengths $\sim 100 \hmpc$.  Even at smaller scales the effective slope
of the power spectrum [or the shape of $\xi(r)$], is poorly constrained, 
with strong variations from sample to sample (see e.g. Branchini \etal 
1994 and references therein). On the other hand, the constraints
on the matter power spectrum from microwave background anisotropy 
measurements as those provided by COBE (Smoot \etal 1992), are on scales 
of $\sim 1000 \hmpc$.  Larger surveys are therefore necessary in order to: a)
establish the slope of $P(k)$ at $\lambda\sim 10-100\hmpc$; b) determine the
scale where $P(k)$ turns over [or equivalently $\xi(r)$ goes negative];
extend the estimates of $P(k)$ from large--scale clustering into a 
regime overlapping with the CMB data.

The LCRS provides interesting hints on $P(k)$ out to wavelengths of $\sim 300
\hmpc$ (Lin 1995), but it is still limited in the accuracy at such wavelengths
due to the survey geometry. The ESP, being a single slice, is going to have 
similar problems in this respect, obviously worsened by the lower sampling.    
Clearly, 3D surveys with typical depth similar to ESP and LCRS, but covering 
wider sky areas are the key for these issues.

\subsection{Connection to Dynamics: Redshift Space Distortions}

The anisotropy introduced by peculiar velocities on the 
observed maps of the galaxy distribution is reflected by 
distortions in the correlation function and power spectrum 
computed in redshift space.  These distortions can be modelled
and  contain a wealth of information about the large--scale 
density field.   Distortions at small
scales are intimately related to the amount of small--scale
power in the power spectrum, i.e. the relative ``temperature'' 
of galaxy pairs.   A low value ($\sim 340$ km s$^{-1}$) for
the pairwise velocity dispersion at $1 \hmpc$ $\sigma_{12}(1)$ 
has been assumed for many years,
on the basis of the result from the CfA1 survey (Davis \& Peebles 1983).
While IRAS galaxies seemed to confirm this figure (Fisher \etal 1994),
more recently, a tendency towards a value closer to ($\sim 600$ km s$^{-1}$) 
has been found in the analysis of larger optical samples (Guzzo \etal
1995; Marzke \etal 1995; see also Zurek, this volume),
but still with large (up to 50\%) fluctuations depending on the volume 
sampled.  That from the full LCRS ($\sim 500$ km s$^{-1}$, Lin
1995), is probably the first estimate sampling a sufficient volume ($\sim 
5\times10^6$ Mpc$^3$) to start seeing convergency of $\sigma_{12}(1)$. 

Distortions of $\xi(r)$ and $P(k)$ in the linear regime (i.e.
at large separations) work in the opposite sense: they are produced
by coherent motions when these are seen parallel to the line of sight 
and have the effect of amplifying clustering.  The amplification is 
approximately expressed as $\beta \simeq {\Omega^{0.6}/ b}$
(Peebles 1980).
Here $b$ is the {\sl bias factor}, the ratio between the rms
fluctuations in the galaxies and in the mass.  By modelling the 
distortions produced on $\xi(r)$ or $P(k)$, under the natural hypothesis
of a fully isotropic underlying clustering process, it is possible
to estimate $\beta$ (e.g. Fisher \etal 1994; Cole \etal 1995).   
However, this is true only if the survey
is geometrically fair, i.e. does not present a dominance of one or
a few structures along some preferred directions.  This is clearly
not the case for $m\le15.5$ surveys containing the Great Wall or 
the Perseus
Pisces chain.  The LCRS finds $\beta = 0.50\pm0.25$ (Lin 1995).
Structures in this and the ESP survey finally seem to be significantly 
smaller than the sample size. However, the errors on $\beta$ can be
improved only with an even larger volume sampled.

\section{The Next Step}

During the next five years at least three major survey projects covering
scales that approach the $\sim 1000\hmpc$ realm will be completed.

\subsection{The Sloan Digital Sky Survey}

Undoubtedly, this represents the largest and most comprehensive 
galaxy survey work ever conceived. It will provide not only 
direct new results of high scientific impact, but also build 
a large photometric and spectroscopic data base to be released to
the scientific community soon after completion of the survey.
The information on the SDSS discussed here come primarily from the
paper by Gunn \& Weinberg (1995, GW95 hereafter), to which the 
reader is referred for details, and from discussions with 
SDSS project members.  Further information can be found on the SDSS www
page ({\tt http://www-sdss.fnal.gov:8000/}).

The project is conducted by a large consortium of U.S. Institutions 
(with the participation of the National Observatory of Japan), 
which has built a dedicated 2.5~m telescope with a corrected 
field of $3^\circ$ in diameter.at Apache Point, New Mexico.  
The auxiliary instrumentation includes a battery of 30 $2048^2$ CCDs 
distributed over six columns,
which will perform imaging in drift scanning mode.  Another 
24 smaller service CCDs will take care of astrometric calibration and 
focusing.  The photometric work will be further assisted by an automatic
0.7~m telescope equipped with a CCD camera and the same filter
set used in the survey camera.  This service instrument will observe 
a network of standard stars to determine extinction corrections and
set accurate zero points for the calibration of both the photometric
and possibly (through narrow--band filters) spectrophotometric data.

For the spectroscopy, two double fibre spectrographs with 320 fibres each
will be used.  Each of them will have a blue and red channel, with independent
cameras and gratings on each arm. 
Interestingly enough, the fibre placing strategy has resorted back to the
classic hand--plugging method on pre--drilled plates. With ten fibre
harnesses, and a total observing--plus--overhead time per plate of about 
a hour, all the plugging work of one night can be done during the day.  
In fact, with two spectrographs available 
that permit to mount the plugged plate on one while the other is observing, 
this solution does not compromise the survey efficiency while allowing a  
significant simplification and reduction of costs.  

\begin{figure}
\vspace{1.75in}
\plotfiddle{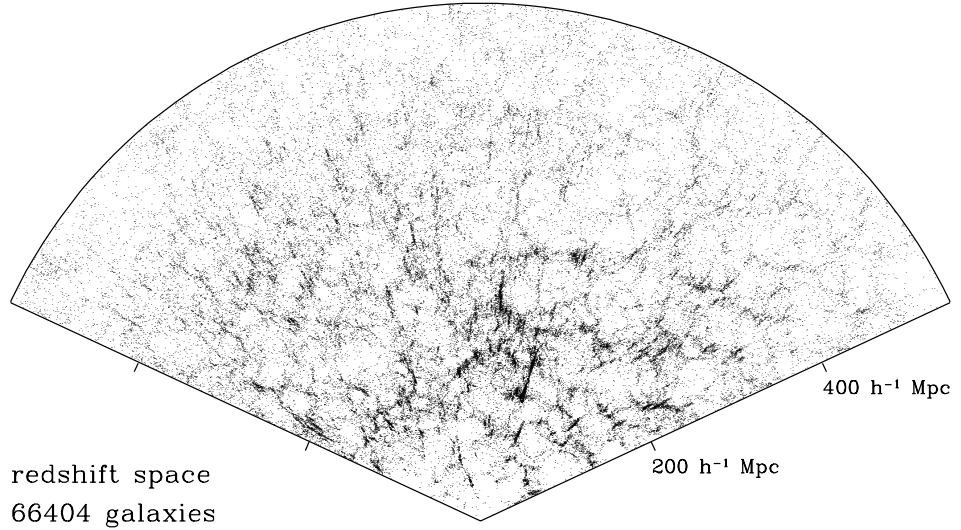}{2truecm}{0}{70}{70}{-216}{-290}
\caption{A $6^\circ$ by $130^\circ$ slice extracted from a large N--body
simulation of a low--density ($\Omega=0.4$, $\Lambda = 0.6$) CDM model
performed by Park \& Gott (from GW95).  The slice is
in redshift space and the selection function has been constructed to reproduce
that expected for the SDSS.  The whole survey will contain about 15 times the 
number of galaxies plotted here.} \label{sdss-cone}
\end{figure}

The essential features of the survey can be summarized as follows:

\begin{itemize}
\item Photometry (S/N=5) in five bands ($u^\prime$,
$g^\prime$, $r^\prime$, $i^\prime$, $z^\prime$), respectively to a depth 
of 22.3, 23.3, 23.1, 22.3, 20.8 over $10^4$ $deg^2$ in the whole North Galactic
Cap.  $200$  deg$^2$ in the South Galactic Cap will be also surveyed to
about 2 magnitudes deeper in each band.

\item This will result in the detection of $\sim 5 \times 10^{7}$ galaxies
and $\sim 10^8$ stars over the $\sim \pi$ steradians of the survey.  From
the latter set of point--like sources, a subset of $\sim 10^6$ 
colour--selected candidate QSOs will be extracted.

\item Medium resolution spectroscopy for the $\sim 10^6$
galaxies brighter than $r^\prime \sim 18$, the $\sim 10^5$ QSOs brighter
than $g\prime \sim 19$ and for specific subsets of stars.

\item The first test year of the SDSS will possibly be started by the 
time these proceedings are published.  The important feature for the community
is that the SDSS consortium is committed to releasing the first two years of 
data within four years from the start of the survey and to opening a public
data archive within two years of its completion, expected within 5 years
from the start.

\end{itemize}


The description of the instrumentation and of the survey strategy
clearly shows how the aims of the SDSS are well beyond the pure 
mapping of the large--scale distribution
of galaxies over the chosen volume.  For a large fraction of the galaxies,
one will have good signal--to--noise photometrically calibrated spectra,
with excellent (3900--9100 \AA) spectral coverage, in addition to the
multi--colour and morphological data from the imaging survey. 
Fig.~\ref{sdss-cone} gives an idea of 
how a $6^\circ$ slice through the SDSS should look like.  It shows a mock
galaxy catalogue extracted from a $600\hmpc$--side simulation by 
 C.~Park \& R.~Gott, containing 54 million particles and started from 
$\Omega=0.4$, $\Lambda = 0.6$ CDM initial conditions (GW95).
The reader might find interesting comparing the qualitative features
of this slice to the real data (with much shallower sampling), of 
fig.\ref{esp-cone}.

\subsection{The 2dF Project}

The 2dF fibre facility at the prime focus of the 
Anglo--Australian Telescope represents
the natural evolution of the UK/Australia leading tradition in fibre
optic instrumentation for wide--field spectroscopy.   2dF is an acronym for 
``2--degree--Field'',
which indicates the main feature of this instrument, namely the capability
of producing corrected images over a $2^\circ$ field of view, to be then 
fed into a powerful 400--fibre automatic positioner. With respect to previous
AAT fibre facilities (FOCAP, AutoFib), this implies a gain of $\sim 9$ in
area coverage, while keeping a similar fibre density on the sky ($\sim 130$
deg$^{-2}$ vs. $\sim 170$ deg$^{-2}$).  The obvious consequence is a similar
gain in the total telescope time required to survey a given area to
the same depth.  Further extending the AutoFib concept, the fibres 
are automatically positioned 
by a flying positioner and magnetically held onto a metal plate placed at
the focal surface.   To minimize time losses due to reconfiguration of the 
fibres, the fibre coupler is doubled, so that one field is configured while
the previous one is being observed.  The 400 fibres are split into two
bundles, which feed two independent spectrographs mounted on the telescope
top ring.  For further details, see Taylor (1995).

The most obvious application of the 2dF facility is large--scale structure
studies.  This is what has been succesfully proposed by a UK/Australia team,
with what is now known as the ``2dF Survey''.   The project concentrates 
on spectroscopy and aims at collecting spectra for 250000 galaxies to around 
$b_J\le 19.7$ (plus a faint extension of further 10000 objects).  
Galaxies will be selected from the APM and EDSGC 
catalogues.  The geometry of the survey will be a combination of a large,
contiguous volume with randomly placed fields.  Two contiguous areas of 
$75^\circ \times 12.5^\circ$ and $65^\circ \times 7.5^\circ$, respectively
in the South and North galactic caps will be fully covered with a
honeycomb of 2dF fields, yielding a total of $\sim 210000$ galaxies.
In addition, 100 randomly--placed fields over the entire southern APM area 
will provide further $\sim 40000$ redshifts over a very large volume.
(It is somewhat amazing to 
consider that these 40000 spectra will be collected in less than 10 nights).
This latter strategy is an interesting feature of the survey, aimed at 
maximising the large--scale signal in the power spectrum whilst keeping
the telescope time within limits acceptable for a general purpose telescope
like the AAT.  Clearly, the SDSS does not
need to resort to such strategical sparse--sampling techniques, since it 
will fully cover (in the North cap), the same area which is sparsely surveyed
by the 100 2dF fields in the South.  

See {\tt http://msowww.anu.edu.au/$\sim$colless/2dF/} for further information.

\subsection{The ESO Redshift Survey of ROSAT Clusters}

The third large--scale project which in the next few years will produce 
a sample exploring
scales of the order of $1000 \hmpc$ involves the use of clusters of
galaxies as tracers of large--scale structure. 

Clusters of galaxies are the largest structures in the Universe to
have clearly separated from the Hubble flow, recollapsing to what can
be considered to some extent a virialized dynamical configuration.
Clusters contain a large quantity of gas at temperatures of 1--10 
keV, which emits in the soft X--ray band with typical luminosities 
around $10^{44}$ erg s$^{-1}$ (see e.g. Sarazin 1986; B\"ohringer 1995).
X--ray clusters can therefore be conveniently used as tracers of the 
large--scale structure of the Universe (see Guzzo 
\etal 1995, and B\"ohringer 1995 for a more detailed discussion).
A redshift survey of clusters of galaxies based on a 
an X--ray wide--area imaging survey represents an 
optimal complement to optically selected galaxy redshift projects.   
A large 3D catalogue of X--ray galaxy clusters
would provide a way to estimate the power spectrum not only through
the usual clustering analysis, but also (at smaller mass scales),
from the luminosity and temperature distribution functions.
The best available X--ray imaging survey from which
starting to construct such a sample is certainly the ROSAT All Sky 
Survey (RASS), performed at the end of 1990 by the ROSAT satellite 
in the 0.1--2.4 keV energy band (Tr\"umper 1992; Voges 1992; 
B\"ohringer 1994).   The first survey processing by the 
so--called Standard Analysis Software System (SASS), yielded a database 
of some 50000 sources, with 4000--5000 of these expected to be clusters 
of galaxies.  Early after completion of SASS, cluster
follow--ups started on specific areas like e.g. the South Galactic Pole
region (Romer \etal 1994; see B\"ohringer 1994 for a review).

Here I would like to describe in detail the ESO redshift survey of
ROSAT clusters of galaxies, an ESO Key--Programme started in 1992 
by a large European team with the aim of measuring the mean redshift of all 
southern clusters with $|b|\ge 20^{\circ}$ and flux larger than 
1.6--2$\times 10^{-12}$ erg s$^{-1}$ cm$^{-2}$ (Guzzo \etal 1995).
The total sample includes $\sim 700$ clusters, and I shall refer 
to it as the KP hereafter.
\smallskip

\noindent{\it Cluster Identification}

Given the 50000 sources detected by SASS, the first step for constructing 
a complete sample is to identify which of these are in fact clusters.  
Clearly, even if it would be highly desirable, direct CCD imaging of all 
SASS sources is presently not feasible, a part from some specific regions
where full indentification is carried out by a specific project at ESO.   
Given the size of the KP, an automatic approach had to be applied.  
Even before starting this identification process, however, one has
to worry on whether the source detection algorithm of SASS has 
actually detected all the sources associated with clusters of galaxies.
In fact, it turns out that nearby, very extended clusters are in fact 
missed by SASS.  This seems to happen only for $z<0.015$, as discussed
in the following.  Once the SASS source list is taken for granted, 
the main identification is then performed by 
cross--correlating the position of the X--ray sources with that of all
optical galaxies with $b_j \le 20.5$ in the southern sky.  
Galaxy positions, magnitudes and
shape parameters are taken from a large reference catalogue, constructed 
in Edinburgh with the COSMOS and SUPERCOSMOS machines from the whole 
southern sky ESO/SRC J survey plates, and handled through a 
collaboration among MPE, ROE--Edinburgh and NRL--Washington.   A source is 
identified as a candidate cluster if it corresponds to a local excess 
in the counts with respect to a certain threshold.
Given the magnitude limit of the galaxy catalogue, the resulting sample 
of clusters has a cutoff at a redshift around $z\sim0.25-0.3$.  
The selection function of the KP, given its flux limit, peaks at redshifts 
between 0.1 and 0.2, so that the redshift cutoff has a limited effect on
the resulting sample.  We also complement the resulting candidate list
with the sources classified as extended in the SASS, but not picked up
by the galaxy excess method.  Although the SASS extension parameter has 
proven not to be a fully reliable estimator (De Grandi, 1996), this
further selection recovers a few clusters missed by the galaxy excess 
method for a number of reasons (e.g. crowding of galaxies).  At the same
time, it picks up some intrinsically very large and luminous objects at 
$z>0.3$, of particular interests even as single objects.  
\smallskip

\noindent {\it Flux Estimation: a Pilot High--Quality Subsample}

\begin{figure}
\vspace{1.0in}
\plotfiddle{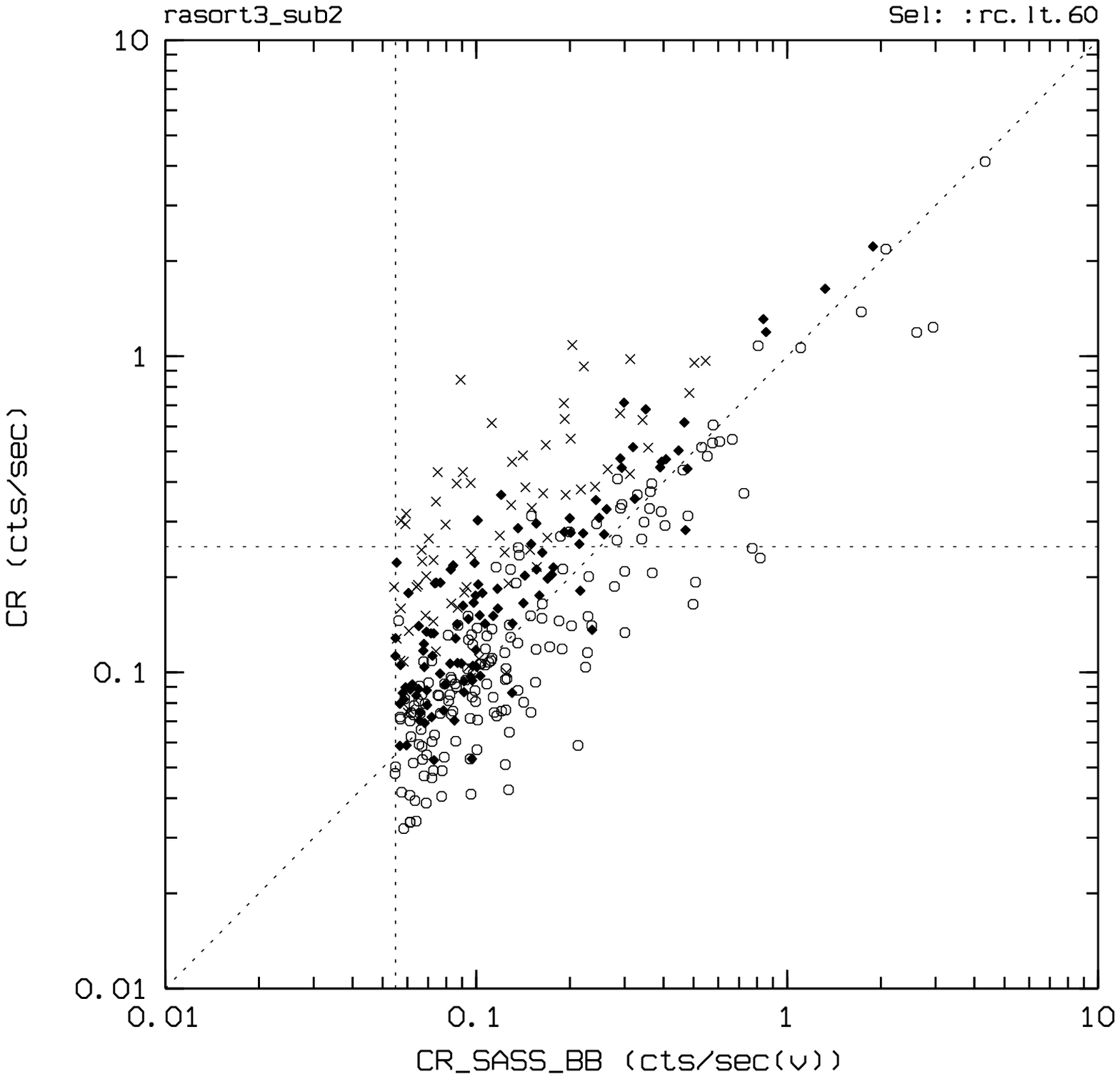}{1.5truecm}{0}{37}{37}{-212}{-55}
\plotfiddle{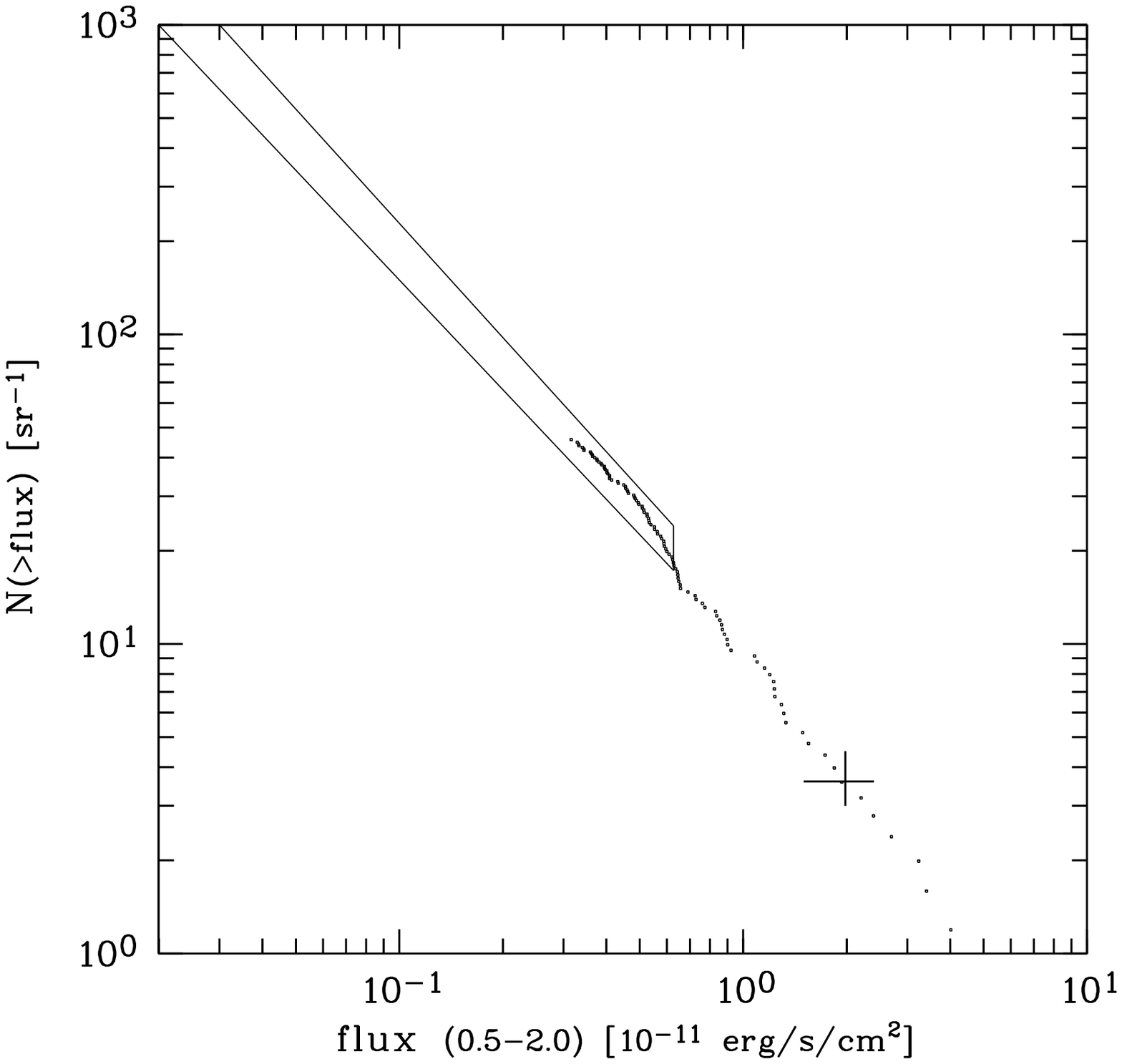}{0truecm}{0}{37}{37}{-15}{-20}
\caption{(From De Grandi 1996). Left panel: Count rates for KP sources 
estimated through the ``steepness ratio'' method as a function of SASS count
rates for the KP candidate cluster sources.  The dashed lines show how an 
initial SASS count limit (CR\_SASS), affects the corresponding true 
completeness limit in the corrected counts (CR).   Open symbols correspond 
to core radii $<$ 60$^{\prime\prime}$, filled symbols to within 
60$^{\prime\prime}$ and 120$^{\prime\prime}$, and crosses to values $>$ 
120$^{\prime\prime}$. Right panel: the observed surface density of clusters 
of the resulting KP X--ray flux--limited pilot sample (dots).  The error cross
is the Piccinotti \etal (1982) result.  The box shows the EMSS
number counts as estimated by Rosati \etal (1995).
} \label{kp-logn}
\end{figure}

Once all the SASS sources candidate for being associated to a cluster
have been identified, a flux--limited sample of clusters can be constructed.
However, as it was recognized early, SASS has the tendency to underestimate 
the count rates for extended sources by about a factor of 2 in the mean, 
(Ebeling, 1993).  A specific effort to understand the survey data and 
re--estimate the X--ray fluxes was then required for assuring the
eventual selection of a complete, truly flux--limited X--ray sample 
for the KP.  Part of this has been done by Sabrina De Grandi for her 
PhD thesis work (1996).  This pilot study concentrates 
on the South Galactic Cap region, and to a final flux limit of
$3\times 10^{-12}$ erg s$^{-1}$ cm$^{-2}$, i.e. about twice as high as that 
of the final KP sample.  The method developed has a number of advantages:
1) it uses the correct RASS point spread function (PSF), 
which has broader wings with respect to the simple Gaussian adopted,
e.g., in the SASS;  2) it allows the 
derivation of a physical extension of the cluster candidates (the core
radius), modelling the profile through a truncated King formula with 
$\beta=2/3$, convolved with the PSF; 3) through 
the same profile, it computes the total counts recovering the missing 
flux from the source wings.   The estimator is robust, being based on
the ratio of two integrals of the King profile (and for this reason called
the {\it Steepness Ratio} method), and has been tested extensively on
sets of well--known sources.
Fig.\ref{kp-logn} (left panel) plots the new estimate of 
the count rates against the corresponding SASS count rates, performed on
the set of candidate KP clusters in the pilot area.  Note the trend, with
the more extended sources being those more underestimated by SASS. 
Also, due to the dispersion of the relation,
any initial cut in the SASS count rates is reflected by a much brighter
cut in the ``true'' counts, if a complete sample has to be obtained.
The same kind of plot done without any initial cut, shows that
with an initial SASS threshold at 0.055 cts s$^{-1}$, the selection 
at corrected count rates $>$ 0.25 cts s$^{-1}$ gives a completeness 
$\geq$  96\%.   The 4\% of missed objects can be shown to correspond to 
very extended sources with angular core radius $\theta_c\geq$ 9.7$^\prime$.  
For $r_c=0.125\hmpc$, this translates into an incompleteness for z $< 0.015$.  

Fig.\ref{kp-logn} (right panel) shows the integral number counts for the
final sample containing 111 cluster, after calculation of the corresponding 
fluxes and sky coverage (De Grandi \etal 1995). 

\smallskip

\noindent {\it Redshift Survey Strategy and Early Results}

The project will use a total of 90 nights, equally distributed among the 
3.6~m, 2.2~m and 1.5~m ESO telescopes.  We typically observe at least 5 
galaxies per cluster, possibly more, to be able to disentangle the 
contamination from galaxy interlopers (e.g. Collins \etal 1995).  
In addition, with $N> 5-10$ redshifts a reasonable estimate of 
the velocity dispersion can be obtained for most of the clusters.
EFOSC at the 3.6~m telescope allows MOS spectroscopy over 
a $\sim 5^\prime \times 5^\prime$ field of view, with 10--20 
redshifts per field typically secured at once on 
moderately distant and/or compact clusters, with 30--40 minute exposures.   
More nearby, looser systems, are instead
surveyed with a long--slit spectrograph at the 2.2~m and 1.5~m telescopes.
At the time these proceedings are published, the survey should have used
more than 60\% of the total telescope time allocated, with more than 250
new candidates observed to be added to other 300 already known cluster
redshifts from the literature or other recent surveys.  Observations 
of the flux--limited pilot subsample of 111 clusters have been virtually 
completed, and a study of its luminosity function is under way.

\begin{figure}
\plotfiddle{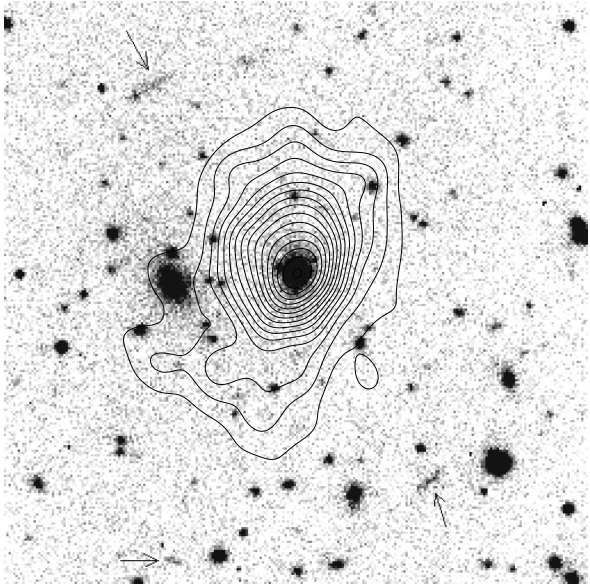}{6truecm}{0}{50}{50}{-120}{0}
\caption{Optical image in the R band of the ultraluminous X--ray cluster 
RXJ 1347.5-1145.  The image side is 1.4 arcmin. The candidate gravitational 
arclets are marked by the arrows, with the two main structures North--East 
and South--West of the central galaxy. The superimposed contours show the 
X--ray emission as seen in a 15760 s exposure with the ROSAT HRI, with 
a resolution of about 5 arcseconds.
} \label{arc}
\end{figure}

As typical of large survey projects, some of the most interesting results
are those which are not expected in the initial plans.  In the case of the KP,
an example is the serendipitous detection of gravitational arcs 
on the short service images collected prior to the MOS spectroscopy
(Edge \etal 1994; Schindler \etal 1995).  The latest discovery
in particular, RXJ 1347.5--1145, has been found to be the most luminous
cluster known in the ROSAT (0.1--2.4 keV) band.   A follow--up campaign 
of observations on this rather interesting object has been started, involving
both ROSAT--HRI and ASCA observations in the X--ray, and ground--based 
NTT plus possibly HST imaging in the optical.  Fig.\ref{arc} shows the 
HRI data superimposed onto the red CCD image on which the arcs were 
originally discovered.

\smallskip

\noindent {\it A Northern Redshift Survey of ROSAT Clusters}

In the Northern hemisphere a redshift survey of ROSAT clusters is also
being performed (see B\"ohringer 1994 for details).  The identification
of clusters in the Northern part of the RASS is complicated by 
the lack of an optical galaxy catalogue comparable in depth and quality
to those obtained in the South from the scanning of the IIIaJ plates.
The survey work is therefore presently limited to a sample selected on
the basis of the SASS extension parameter, a criterion that at present
cannot guarantee the selection of all the true RASS X--ray clusters 
within a certain region.  It will be therefore necessary to complement 
the sample with a more general cluster identification work as 
soon as a good galaxy data base (e.g. the SDSS) becomes available 
in the Northern sky.  There is a strong case for this, since in this
way the Southern and Northern surveys could be homogenized and 
combined, providing a total sample with about 1500 clusters at 
a typical depth of $\sim 600$ $\hmpc$.  This would be
more than invaluable for all those cosmological investigations
requiring all--sky coverage, as e.g. the study of dipoles and peculiar
motions.   

See also {\tt http://www.merate.mi.astro.it/$\sim$guzzo/KP.html} for
further information.

\begin{figure}
\vspace{1.0in}
\plotfiddle{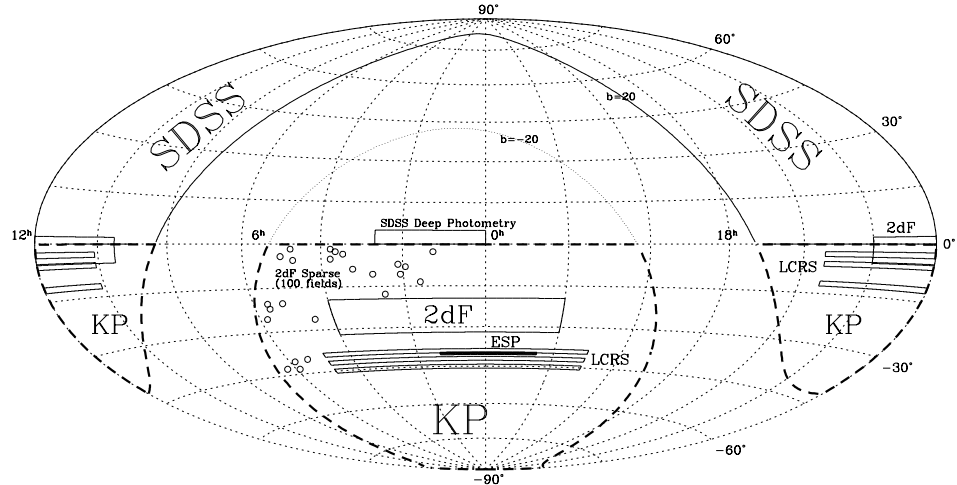}{2truecm}{0}{50}{50}{-200}{-50}
\caption{An Aitoff RA--DEC all--sky projection showing the location
of the major galaxy and cluster survey projects discussed in the text.   
} \label{allsky}
\end{figure}

\section{Conclusions}

I believe a good conclusion here is to say that we seem to live in a very 
fortunate and exciting epoch, with all these very promising projects 
due to finish by the end of the Millenium.   In a few years most of the 
sky will be covered spectroscopically down to a typical depth of $\sim 600 
\hmpc$ and the amount of data and information at our disposal will be 
increased 
by two orders of magnitude.   I thought the best way to have the feeling
of how the sky will be attacked during the forthcoming years was to try and
condense all I have talked about in this review into a single picture.
After some headache with SM, the result is shown in fig.\ref{allsky}.


\acknowledgments

I would like to thank all my collaborators in the various projects
discussed here for the privilege of discussing our common work.  
I thank G. E. Zucca for useful discussions, D. Weinberg and M. Strauss
for providing data and information on the SDSS, and C. Collins for
providing information on the 2dF project.  I am grateful to E. 
Molinari for help with the figures and to M. Strauss for the use 
of his SM macros.



%
%

%

\end{document}